\documentclass[11pt]{article}

\usepackage[utf8]{inputenc}
\usepackage{charter}
\usepackage{fullpage}
\usepackage{hyperref}
\usepackage{graphicx}
\usepackage{lipsum}
\usepackage[sort&compress,numbers]{natbib}
\usepackage{hyperref}
\hypersetup{hidelinks}
\usepackage[total={6.5in,9in},top=1in,headsep=0.1in,headheight=1in]{geometry}
\usepackage{enumitem,amssymb}
\usepackage[usenames,dvipsnames]{xcolor}
\usepackage{xfrac}
\usepackage{bm}
\usepackage{array}
\usepackage{comment}
\usepackage{makecell}
\usepackage{ulem}
\usepackage{epstopdf}

\newcommand\snowmass{
\begin{center}
  \rule[-0.2in]{\hsize}{0.01in}\\
  \rule{\hsize}{0.01in}\\
  \vskip 0.1in
  Submitted to the Proceedings of the US Community Study\\ 
  on the Future of Particle Physics (Snowmass 2021)\\
  \rule{\hsize}{0.01in}\\
  \rule[+0.2in]{\hsize}{0.01in}\\[-2em]
\end{center}
}

\usepackage[firstpage=true]{background}
\backgroundsetup{contents={\parbox{6.5in}{\snowmass}}, scale=1,placement=top,opacity=1,color=black,position={3.25in,1.2in}}

\usepackage{fancyhdr}
\fancypagestyle{plain}{%
  \fancyhf{}%
  \fancyhead[C]{}
  \fancyfoot[C]{\thepage}
}

\fancypagestyle{empty}{%
  \fancyhf{}%
  \fancyhead[C]{{\it Cosmology with mm-wave line intensity mapping}}
  \fancyfoot[C]{\thepage}
}
\title{Snowmass 2021 Cosmic Frontier White Paper:\\ Cosmology with Millimeter-Wave Line Intensity Mapping}
\date{}

\usepackage{authblk}

\author[1,2]{Kirit~S.~Karkare}
\author[3]{Azadeh Moradinezhad Dizgah}
\author[4]{Garrett~K.~Keating}
\author[5]{\\ Patrick Breysse}
\author[6,7]{Dongwoo~T.~Chung}
\author[\space]{\\ for the Snowmass 2021 Cosmic Frontier 5 Topical Group\vspace{.2in}}
\author[8]{\\ \textit{Endorsers:} James Aguirre}
\author[9,10]{Zeeshan Ahmed}
\author[2]{Adam Anderson}
\author[11]{Pete~S.~Barry}
\author[12]{Ritoban Basu Thakur}
\author[2,1]{Bradford Benson}
\author[13]{Jos\'e Luis Bernal}
\author[9,10]{Federico Bianchini}
\author[14]{Simeon Bird}
\author[15,12]{C.~Matt Bradford}
\author[1,11]{John~E.~Carlstrom}
\author[16]{Emanuele Castorina}
\author[17]{Andrea Caputo}
\author[11,1]{Clarence Chang}
\author[15,12]{Tzu-Ching Chang}
\author[12]{Yun-Ting Cheng}
\author[13]{Cyril Creque-Sarbinowski}
\author[18]{Abigail~T.~Crites}
\author[15,12]{Oliver Dor\'e}
\author[19,20,21]{Jacques Delabrouille}
\author[21]{Simone Ferraro}
\author[22]{Jeffrey Filippini}
\author[23]{Yan Gong}
\author[24]{Ely Kovetz}
\author[25]{Guilaine Lagache}
\author[8]{Adam Lidz}
\author[5]{Abhishek~S.~Maniyar}
\author[26]{Daniel~P.~Marrone}
\author[1]{Jeff McMahon}
\author[11]{Zhaodi Pan}
\author[21]{Emmanuel Schaan}
\author[1]{Erik Shirokoff}
\author[2]{Sara~M.~Simon}
\author[18]{Gordon Stacey}
\author[27]{Eric~R.~Switzer}
\author[28]{Peter Timbie}
\author[22]{Joaquin Vieira}
\author[11]{Gensheng Wang}
\author[10]{W.~L.~Kimmy Wu}
\author[29]{Michael Zemcov \vspace{.2in}}

\affil[1]{Kavli Institute for Cosmological Physics, University of Chicago, Chicago, IL 60637, USA}
\affil[2]{Fermi National Accelerator Laboratory, Batavia, IL 60510, USA}
\affil[3]{D\'epartement de Physique Th\'eorique, Universit\'e de Gen\`eve, 1211 Gen\`eva 4, Switzerland}
\affil[4]{Center for Astrophysics, Harvard \& Smithsonian, Cambridge, MA 02138, USA}
\affil[5]{Center for Cosmology and Particle Physics, New York University, New York, NY 10003, USA}
\affil[6]{Canadian Institute for Theoretical Astrophysics, University of Toronto, Toronto, ON M5S 3H8, Canada}
\affil[7]{Dunlap Institute for Astronomy and Astrophysics, University of Toronto, Toronto, ON M5S 3H4, Canada}

\affil[8]{University of Pennsylvania, Philadelphia, PA 19104, USA}
\affil[9]{Kavli Institute for Particle Astrophysics and Cosmology, Stanford University, Stanford, CA 94305, USA}
\affil[10]{SLAC National Accelerator Laboratory, Menlo Park, CA 94025, USA}
\affil[11]{Argonne National Laboratory, Lemont, IL 60439, USA}
\affil[12]{California Institute of Technology, Pasadena, CA 91125, USA}
\affil[13]{Johns Hopkins University, Baltimore, MD 21218, USA}
\affil[14]{University of California, Riverside, Riverside, CA 92521, USA}
\affil[15]{Jet Propulsion Laboratory, Pasadena, CA 91109, USA}
\affil[16]{University of Milan, 20122 Milano MI, Italy}
\affil[17]{Tel Aviv University, Ramat Aviv 69978, Israel}
\affil[18]{Cornell University, Ithaca, NY 14850, USA}
\affil[19]{Centre Pierre Bin\'etruy International Research Laboratory, CNRS, 75016 Paris, France}
\affil[20]{University of California, Berkeley, Berkeley, CA 94720, USA}
\affil[21]{Lawrence Berkeley National Laboratory, Berkeley, CA 94720, USA}
\affil[22]{University of Illinois, Urbana-Champaign, Champaign, IL 61801, USA}
\affil[23]{National Astronomical Observatories, Chinese Academy of Sciences, Beijing, China}
\affil[24]{Ben-Gurion University of the Negev, Be'er Sheva, Israel}
\affil[25]{Aix Marseille University, CNRS, CNES, LAM, Marseille, France}
\affil[26]{University of Arizona, Tucson, AZ 85721, USA}
\affil[27]{Goddard Space Flight Center, Greenbelt, MD 20771, USA}
\affil[28]{University of Wisconsin, Madison, Madison, WI 53715, USA}
\affil[29]{Rochester Institute of Technology, Rochester, NY 14623, USA}

\begin{document}

\maketitle

\clearpage
\begin{abstract}
Next-generation tests of fundamental physics and cosmology using large scale structure require measurements over large volumes of the Universe, including high redshifts inaccessible to present-day surveys. Line intensity mapping, an emerging technique that detects the integrated emission of atomic and molecular lines without resolving sources, can efficiently map cosmic structure over a wide range of redshifts. Observations at millimeter wavelengths detect far-IR emission lines such as CO/[CII], and take advantage of observational and analysis techniques developed by CMB experiments. These measurements can provide constraints with unprecedented precision on the physics of inflation, neutrino masses, light relativistic species, dark energy and modified gravity, and dark matter, among many other science goals. In this white paper we forecast the sensitivity requirements for future ground-based mm-wave intensity mapping experiments to enable transformational cosmological constraints. We outline a staged experimental program to steadily improve sensitivity, and describe the necessary investments in developing detector technology and analysis techniques.
\end{abstract}

\section{Introduction} \label{sec:intro}

Over the past few decades, thanks to the increasing volume and precision of cosmological surveys---in particular measurements of the cosmic microwave background (CMB) and large-scale structure (LSS)---we have established a high-precision, ``concordance'' model of cosmology: the $\Lambda$CDM paradigm. Nevertheless, we are still left with fundamental open questions. What was the mechanism that set the seeds of cosmic structure in the very early universe? What is the nature of dark energy (DE) and dark matter (DM)? What are the properties of neutrinos and other possible light relics? High-precision maps of the three-dimensional LSS provide a rich trove of information about the origin, evolution and composition of the Universe, probing deviations from the standard cosmological model and potentially hinting at new physics. The line intensity mapping (LIM) technique offers a means to map LSS over a wide range of scales and across many redshift epochs, complementing upcoming wide-field galaxy surveys at redshifts  $z\leq 2$, and providing a spectroscopic probe of LSS at $z>2$.

\begin{figure}[bp!]
\centering
\includegraphics[width=0.9\textwidth]{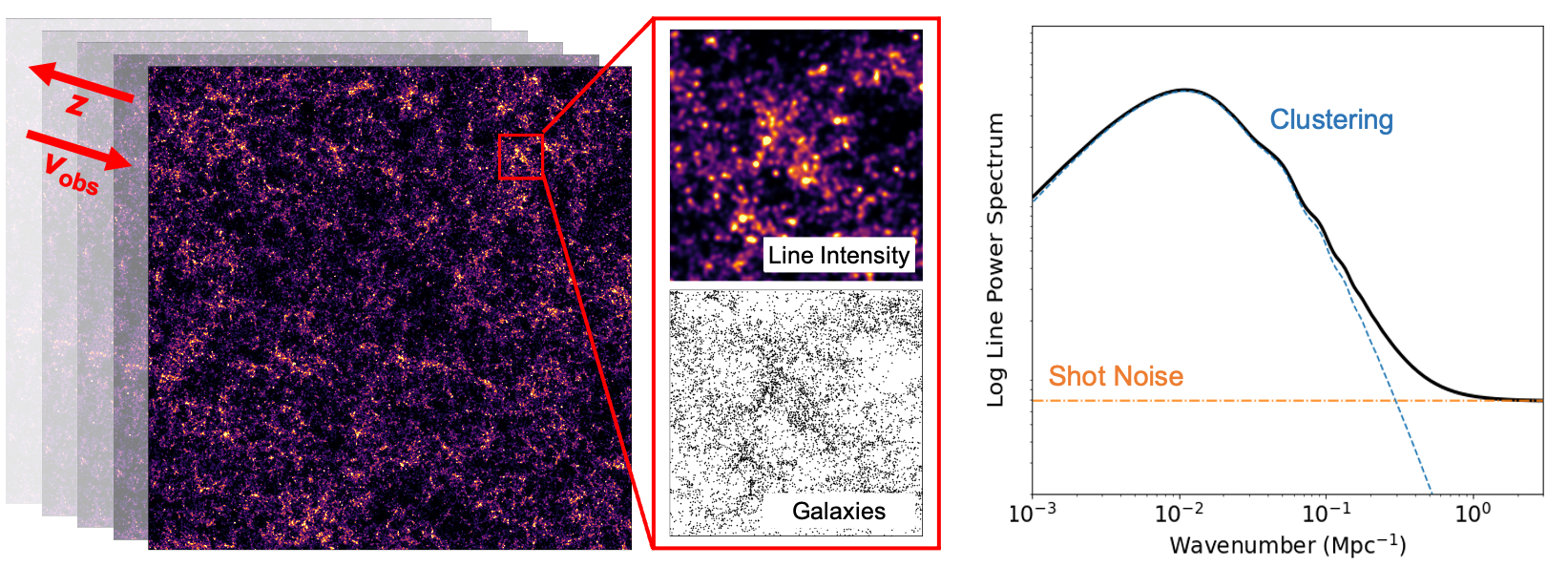}
\caption{\textit{Left}: Schematic image of a 100 deg$^2$ line intensity map. Individual layers indicate the three-dimensional information obtained by mapping at different frequencies $\nu_{\rm obs}$. \textit{Center}: Zoom-in on a small segment of an individual frequency channel, showing the correspondence between the large-scale structure seen in the underlying galaxy population and the lower-resolution intensity map. \textit{Right}: Schematic of a LIM power spectrum (solid black), showing the large-scale clustering component (dashed blue) driven by the dark matter distribution and the shot noise component (dot-dashed orange) due to the Poisson statistics of individual objects.}
\label{fig:overview}
\end{figure}

In contrast to typical galaxy surveys, which directly image large numbers of resolved galaxies, LIM measures the aggregate emission of all galaxies in a chosen spectral line \cite{kovetz2017}. Spatial fluctuations in the resulting intensity field and spectral emission information can be combined to generate a smoothed three-dimensional map of the galaxy distribution. As illustrated in Figure~\ref{fig:overview}, the clustering component traces the underlying DM while the shot noise contribution arises from the discrete nature of the sources. Emission lines originating from different phases of the interstellar medium can be used individually and in synergy with each other to constrain cosmology and astrophysics \cite{Fonseca:2016qqw,Sun2019,Schaan:2021gzb}. \textbf{Since the LIM technique does not require resolving individual galaxies, it is an efficient probe of LSS over large comoving volumes, enabling measurements over large scales and over long redshift lever arms, including the high-redshift regime}. The promise of LIM surveys for testing fundamental cosmology is therefore multifold: 
\begin{enumerate}
\item Wide redshift coverage mitigates degeneracies between cosmological parameters due to different degeneracy directions at low and high redshifts. 
\item Measuring very large-scale modes allows extraction of key information about physics at the scale of the cosmological horizon, including imprints of multiple degrees of freedom during inflation and general relativistic effects that probe the theory of gravity.  
\item Probing higher redshifts provides access to an order of magnitude more linear and semi-linear density modes, which are less affected by nonlinearities and astrophysical processes that decorrelate the modes from the initial conditions, and can be described accurately with perturbative theoretical models; see the dedicated Snowmass 2021 white paper on cosmology at high redshift for further details \cite{Snowmass2021:largeN}. 
\end{enumerate}

\begin{figure}[t]
    \centering
    \includegraphics[width=\textwidth]{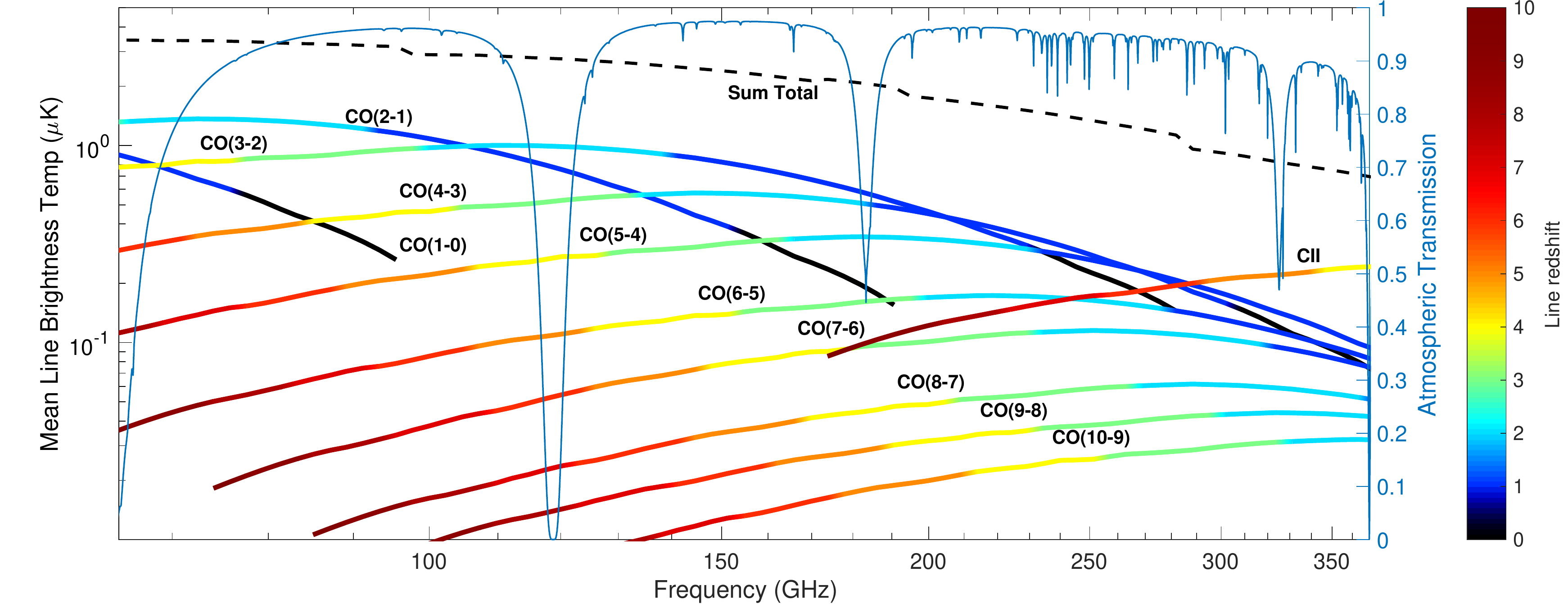}
    \caption{ Spectral lines detectable by a ground-based mm-wavelength survey (reproduced from Ref.~\cite{MoradinezhadDizgah:2021upg}). Shown above are model predictions for the brightness temperatures of individual lines, their sum (dashed black), and the median atmospheric transmission at the South Pole (light thin blue). Colors correspond to observed redshifts of the lines.}
    \label{fig:atmosphere}
\end{figure}

LIM at millimeter wavelengths has garnered significant experimental interest in recent years. These measurements target atomic and molecular lines with rest-frame wavelengths in the far-IR. Lines such as the CO(J$\rightarrow$J-1) rotational transitions and the [CII] ionized carbon fine structure line are bright in dusty, star-forming galaxies that are abundant in the early Universe, and have been observed in individual sources out to the epoch of reionization \cite{pentericci2016, Combes2018}. Observations from cm to THz frequencies are capable of detecting emission from redshifts $0 < z < 10$ using a combination of lines. Figure~\ref{fig:atmosphere} shows the mean brightness temperatures of  spectral lines detectable by mm-wavelength ground-based surveys, in addition to representative atmospheric transmission. These observations operate in a region of frequency space with much less Galactic foreground contamination than LIM efforts targeting the 21 cm line \cite{switzer2019}, and build on the extensive heritage of CMB observations at similar frequencies.  Current  experiments, including CCAT-p \cite{CCAT2021}, COMAP \cite{Cleary2021}, CONCERTO \cite{Lagache2018}, EXCLAIM \cite{Cataldo2021}, mmIME \cite{Keating2020}, SPT-SLIM \cite{Karkare2021}, TIM \cite{Vieira2020}, and TIME \cite{Crites2014}, are demonstrating the technologies and analysis techniques to probe cosmology, but so far lack the necessary scale and sensitivity to place competitive cosmological constraints.  

The goal of this white paper is to illustrate that next-generation mm-wave LIM experiments are capable of testing fundamental physics and cosmology, particularly at higher redshifts and over larger volumes than possible with current probes of LSS. After describing the motivating science cases, we forecast the measurement requirements needed to cross critical target thresholds. We outline a staged experimental program that reaches these thresholds by gradually increasing the sensitivity and volume of LIM observations. We then describe the needed investments in technology to improve experimental sensitivity, and summarize the challenges in extracting cosmology from LIM measurements and the investments in analysis techniques that would address them.

\section{Science Summary} \label{sec:science}

Wide-field LIM surveys would advance our understanding of open fundamental questions in cosmology, and in several cases cross physically-motivated theoretical thresholds. Measuring clustering statistics of line intensity fluctuations over the redshift range $0<z<10$ would allow us to: 
\begin{enumerate}
    \item {\bf Decipher the origin of cosmic structure:} 
    The simplest models of single-field, slow-roll inflation predict primordial fluctuations with a nearly scale-invariant power spectrum and Gaussian distribution, in agreement with  CMB and LSS observations \cite{Planck:2018jri}. Deviations from either prediction, e.g., features in the primordial power spectrum \cite{Chluba:2015bqa,Slosar:2019gvt} or primordial non-Gaussianity (PNG) \cite{Meerburg:2019qqi}, would probe fundamental physics at the highest energy scales and shed light on the mechanism that set the seeds of cosmic structure; see the dedicated Snowmass 2021 white paper on Inflation \cite{Snowmass2021:Inflation}. High-precision measurements of the line power spectrum can significantly improve current bounds on inflaton features \cite{Planck:2018jri,Beutler:2019ojk,Ballardini:2022wzu} by detecting, e.g., sinusoidal modulation of the power spectrum \cite{Sailer:2021yzm} with enhanced sensitivity to lower amplitudes and a broader range of feature frequencies. The amplitude of the local-type PNG, $f_{\rm NL}^{\rm loc}$, can be constrained by measuring the scale-dependent line bias on large scales \cite{Dalal:2007cu}. Furthermore, measurement of the line bispectrum improves limits on the local shape and constrains other types of PNG, such as the equilateral \cite{Creminelli:2003iq,Alishahiha:2004eh,Cheung:2007st}, orthogonal \cite{Senatore:2009gt}, and folded \cite{Chen:2006nt,Holman:2007na} shapes (which do not leave distinguishable imprints on the power spectrum \cite{Karagiannis:2019jjx}). Detection of local-type PNG is considered a ``smoking gun'' for multi-field inflation, which generically predicts $f_{\rm NL}^{\rm loc} = {\mathcal O}(1)$. Therefore, {\bf crossing the critical threshold of $\bm{\sigma(f_{\rm NL}^{\rm loc}) \sim 1}$ would rule out single-field models of inflation}, which obey the consistency relation enforcing $f_{\rm NL}^{\rm loc} \ll 1$ \cite{Maldacena:2002vr,Creminelli:2004yq}. Reaching similar thresholds on other shapes is also desirable:\ {\bf detecting $\bm{f_{\rm NL}^{\rm eq} \geq 1}$ would imply that inflation is a strongly-coupled phenomenon} \cite{Baumann:2011su,Baumann:2014cja} and  considerably limit inflation model space. CMB data from Planck provide the current tightest constraints \cite{Planck:2019kim}, but recent results from galaxy surveys \cite{Rezaie:2021voi,Cabass:2022wjy,DAmico:2022gki} point to the potential of LSS to exceed the current CMB limits of $f_{\rm NL}^{\rm loc} = -0.9 \pm 5.1, \ f_{\rm NL}^{\rm eq} = -26 \pm 47, f_{\rm NL}^{\rm orth} = -38 \pm 24$. \textbf{\textit{Advantages of LIM}}: First, LIM measurements of large comoving volumes compared to galaxy surveys would access very large-scale modes, which are crucial in constraining $f_{\rm NL}^{\rm loc}$ using scale-dependent bias. Second, since the signal-to-noise ratio for PNG in the bispectrum sharply increases as a function of the smallest scale \cite{Sefusatti:2011gt}, probing $z>2$ and accessing a large number of linear and semi-linear modes make LIM surveys particularly powerful compared to lower-redshift surveys. Third, using cross-correlations between multiple lines observable in a LIM survey in addition to auto spectra can further improve constraints by mitigating cosmic variance \cite{Seljak:2008xr}. 
    
    \item {\bf Measure the sum of neutrino masses:} When neutrinos become non-relativistic, their masses contribute to the matter content of the universe and modify the expansion history and growth of structure \cite{Lesgourgues:2013sjj}. They can thus be constrained by measurements of line clustering statistics such as the power spectrum. The mass splitting measured by neutrino oscillation experiments sets a minimum for the sum of neutrino masses, $M_\nu = 0.059 \ {\rm eV}$ in the normal and $M_\nu = 0.10 \ {\rm eV}$ in the inverted mass hierarchies \cite{deSalas:2017kay,Esteban:2018azc}. \textbf{This sets a critical threshold of $\bm{\sigma(M_{\nu}) \sim 0.012 \ {\rm eV}}$ for a $\bf{5\sigma}$ detection of the minimum neutrino mass} \cite{Dvorkin:2019jgs}. Improving on the current best constraints of $\sigma(M_\nu)\leq 0.12 \ {\rm eV}$ from CMB and LSS data \cite{Planck:2018vyg, Ivanov:2019hqk} would indirectly distinguish between the two mass orderings, with implications for determining the nature of neutrinos (Dirac vs.\ Majorana), understanding mixing in the lepton sector, and discriminating between flavor models \cite{DeSalas:2018rby}; see the dedicated Snowmass 2021 white paper on synergies between cosmological and laboratory probes of neutrinos \cite{Snowmass2021:NeutrinoSynergy}. \textbf{\textit{Advantages of LIM}}: By mapping LSS over a long redshift lever arm, LIM surveys capture the redshift evolution of the scale-dependent suppression of the matter power spectrum and growth rate due to neutrino masses. Furthermore, wide redshift coverage breaks parameter degeneracies in the CMB and low-redshift LSS probes \cite{Yu:2018tem}, most notably between $M_\nu$ and the large-scale amplitude of the matter power spectrum, optical depth, and DE equation of state \cite{Hannestad:2005gj,Liu:2015txa,Allison:2015qca}.
  
    \item {\bf Search for new light particles:} The Standard Model (SM) of particle physics predicts three species of massless neutrinos contributing to the radiation content of the Universe at early times, corresponding to $N^{\rm SM}_{\rm eff} = 3.046$. Changing $N_{\rm eff}$ modifies the amplitude of the matter power spectrum and the phase and amplitude of the baryon acoustic oscillations \cite{Baumann:2017lmt}, and thus can be constrained by measuring the power spectra of biased tracers of DM. Detection of excess light relic abundance offers discovery space for new physics, as many extensions to the SM predict extra light relics.  The spin states and decoupling temperatures determine the minimum excess $\Delta N_{\rm eff}$ values: $\Delta N_{\rm eff} > 0.027$ for a single scalar, $0.047$ for a Weyl fermion, and $0.054$ for a vector boson \cite{Brust:2013ova,Chacko:2015noa}. Therefore, \textbf{reaching $\bf{\sigma(N_{\rm eff}) \lesssim 0.02-0.03}$ would constrain excess light relics with non-zero (zero) spin at $\bf{2\sigma}$ ($\bf{1\sigma}$)}, probing a wide range of models---including those with axions, axion-like particles, thermal populations of gravitinos, and dark photons \cite{CMB-S4:2016ple}; see the dedicated Snowmass 2021 white paper on light relics \cite{Snowmass2021:LightRelics}. \textbf{\textit{Advantages of LIM}}: By probing extremely large comoving volumes, LIM surveys could significantly improve on current limits from CMB and LSS data, $\sigma(N_{\rm eff}) \simeq 0.17$ \cite{Planck:2018vyg}. Since LIM is not affected by the degeneracy between $N_{\rm eff}$ and the primordial Helium fraction, $Y_{\rm He}$---one of the limiting factors in CMB constraints on light relics---these measurements will be highly complementary to CMB-S4 \cite{Abazajian:2019eic}. 
    
    \item{\bf Test the physics of dark energy and gravity:} 
    Although the current accelerated expansion is well-modeled by a cosmological constant which begins to dominate the cosmic energy density at late times, $z \lesssim 2$, it unclear whether this is the correct model of DE, or if the acceleration is due to modifications to General Relativity.
    ``Tracking''-type behavior, where DE behaves like matter at high redshift with equation of state $w\simeq 0$, and then transitions to a cosmological constant with $w \simeq -1$ at low redshift, is a generic prediction of scalar-tensor theories with second-order equations of motion (Horndeski theories) \cite{Raveri:2017qvt, Traykova:2021hbr}. Another class of models, which has received attention as a solution to the Hubble tension  \cite{Karwal:2016vyq,Niedermann:2019olb,Braglia:2020bym}, assumes an early dark energy (EDE) component that behaves like a cosmological constant before some critical pre-recombination redshift $z_c$, and afterwards dilutes faster than radiation. LSS surveys can test these models by precisely characterizing the expansion history during both the matter- and DE-dominated eras \cite{Bull:2020cpe}.  At the same time, LSS probes models of gravity by tracing their imprints on the growth of structure \cite{Jain:2010ka,Joyce:2016vqv}; while  modifications to gravity and dynamic DE are largely indistinguishable in the expansion history, their effects are distinct in the fluctuations probed by clustering statistics \cite{Dodelson:2016wal}. \textbf{\textit{Advantages of LIM}}: High-precision LIM measurements would significantly improve current constraints on dynamical and EDE models due to wide redshift access. \cite{Ivanov:2020ril,Hill:2020osr,Smith:2020rxx}. Measurements at high redshift probe the EDE fraction out to very high pre-recombination redshifts \cite{Sailer:2021yzm}, and constrain EDE where its effect on the matter power spectrum is more prominent since nonlinearities reduce the difference between EDE and $\Lambda$CDM \cite{Klypin:2020tud}.   
    
    \item {\bf Explore dark matter candidates:} High-precision measurements of line clustering statistics constrain various dark matter scenarios, including those that act like an additional contribution to $N_{\rm eff}$ \cite{Green:2017ybv} and those that leave features in the clustering statistics of DM tracers, such as self-interacting DM \cite{Tulin:2017ara}, ultra-light axion DM \cite{Hlozek:2014lca}, and models with DM-baryon scattering \cite{Dvorkin:2013cea}; see the dedicated Snowmass 2021 white papers on dark matter constraints from cosmic surveys \cite{Snowmass2021:DMhalosWP,Snowmass2021:DMsimsWP,Snowmass2021:DMfacilitiesWP}. \textbf{\textit{Advantages of LIM}}: By probing the faintest galaxies, which are not detected by galaxy surveys, LIM can uniquely constrain DM models that suppress small-scale structure \cite{Bauer:2020zsj}.  LIM surveys can also be sensitive to radiative decays/annihilations of a broad range of beyond-the-Standard-Model particles which decay to photons at a specific frequency. If the density of these exotic species traces the DM, they will appear in LIM data as an emission line with no known atomic or molecular frequency. Models that can be probed using this signature include axion or sterile neutrino dark matter which decays to photons \cite{CrequeSarbinowski2018,Bernal2021}, or an enhanced decay rate of the cosmic neutrino background \cite{Bernal2021a}. 
    
\end{enumerate}

Beyond the fundamental physics goals outlined above, a wide-field high-redshift LIM survey would also be a valuable tool for a broader range of astrophysical studies; see e.g., \cite{kovetz2017, Kovetz2019}. LIM surveys provide a unique window into  the aggregate properties of faint, distant galaxies unobservable by traditional surveys, probing galaxy growth and evolution \cite{Keating2016,Yang2019,Keating2020,Breysse2021,Cleary2021,Anderson2022} as well as the process of cosmic reionization \cite{Lidz2011}.

\section{Measurement Requirements and Staged Experimental Program}

Focusing on a subset of science goals described above, in this section we forecast the experimental sensitivity that LIM surveys targeting CO/[CII] require to cross the critical thresholds for $f_{\rm NL}^{\rm loc}, M_\nu$, and $ N_{\rm eff}$. We also use the Chevallier-Polarski-Linder parametrization of the DE equation of state ($w_0, w_a$) to illustrate LIM's potential to constrain dynamic DE models \cite{Chevallier:2000qy,Linder:2002et}. 

Rather than focusing on a single survey design, we perform Fisher forecasts to determine the optimal design of ground-based mm-wave LIM surveys in achieving the best constraints on parameters of interest by measuring CO/[CII]. We present constraints as a function of \textit{spectrometer-hours}: the product of spectrometer count and integration time, which is a reasonable proxy for sensitivity and the ``effort level'' of an experiment. Based on these results, we then outline a staged experimental approach that builds up sensitivity, enabled by technical advances discussed in Section~\ref{sec:invest}.

We use only the line power spectra in our forecasts. Exploiting information in the observed intensity field beyond the power spectrum, e.g., higher-order statistics, one-point statistics, and synergies between LIM and future CMB and galaxy surveys \cite{Snowmass2021:jointProbes} would not only improve constraints, but also offer a means to overcome degeneracies with nuisance astrophysical parameters, and to improve mitigation of systematics and foregrounds. 

\subsection{Sensitivity required to cross critical thresholds}

Our forecasts consider power spectrum measurements of several CO rotational lines (from J=2-1 to J=6-5) and the [CII] fine-structure line combined with Planck primary CMB data \cite{Planck:2018vyg}. As the best- and worst-case scenarios, we show constraints with and without interloper lines as a source of noise in our analysis \cite{Lidz:2016,Cheng:2016}. We use a linear model of the line power spectrum, and assume that the line biases and mean brightness temperatures are well-constrained. We choose a conservative bound on the small-scale cutoff ($k_{\rm max}$) to stay in the regime where the linear model of the line power spectrum is accurate to $\sim 20\%$ \cite{MoradinezhadDizgah:2021dei}. The large-scale cutoffs---most relevant for $f_{\rm NL}^{\rm loc}$---are set by the angular extent of the survey and the atmospheric transmission as a function of frequency in the angular and line-of-sight directions, respectively. For each line, we split the survey into redshift bins with mean redshifts $z_i$ and widths of 0.1 dex to account for the cosmic evolution of the line-emitting population, as well as variations in noise levels caused by the atmospheric transmission. The constraints are obtained using a unified Fisher Forecast pipeline described in Ref.~\cite{MoradinezhadDizgah:2021upg}, where further details about the signal and interloper modeling, instrument noise, and forecasting assumptions are outlined. 

The instantaneous sensitivity of an instrument is determined by spectrometer count. Each ``spectrometer'' consists of a single spatial pixel that fully samples the 80--310 GHz range with frequency channels of width $\Delta \nu = \nu / R$; we assume resolving power $R =300$. Millimeter-wave detectors are now routinely fabricated with device noise that is subdominant to the atmospheric background at excellent sites \cite{baselmans2017}---even at the $R \sim 100-1000$ spectral resolution needed for LIM science---so we assume that the instrument noise is dominated by thermal emission from the atmosphere.

\begin{figure}[htbp!]
    \centering
    \includegraphics[width=0.45 \textwidth]{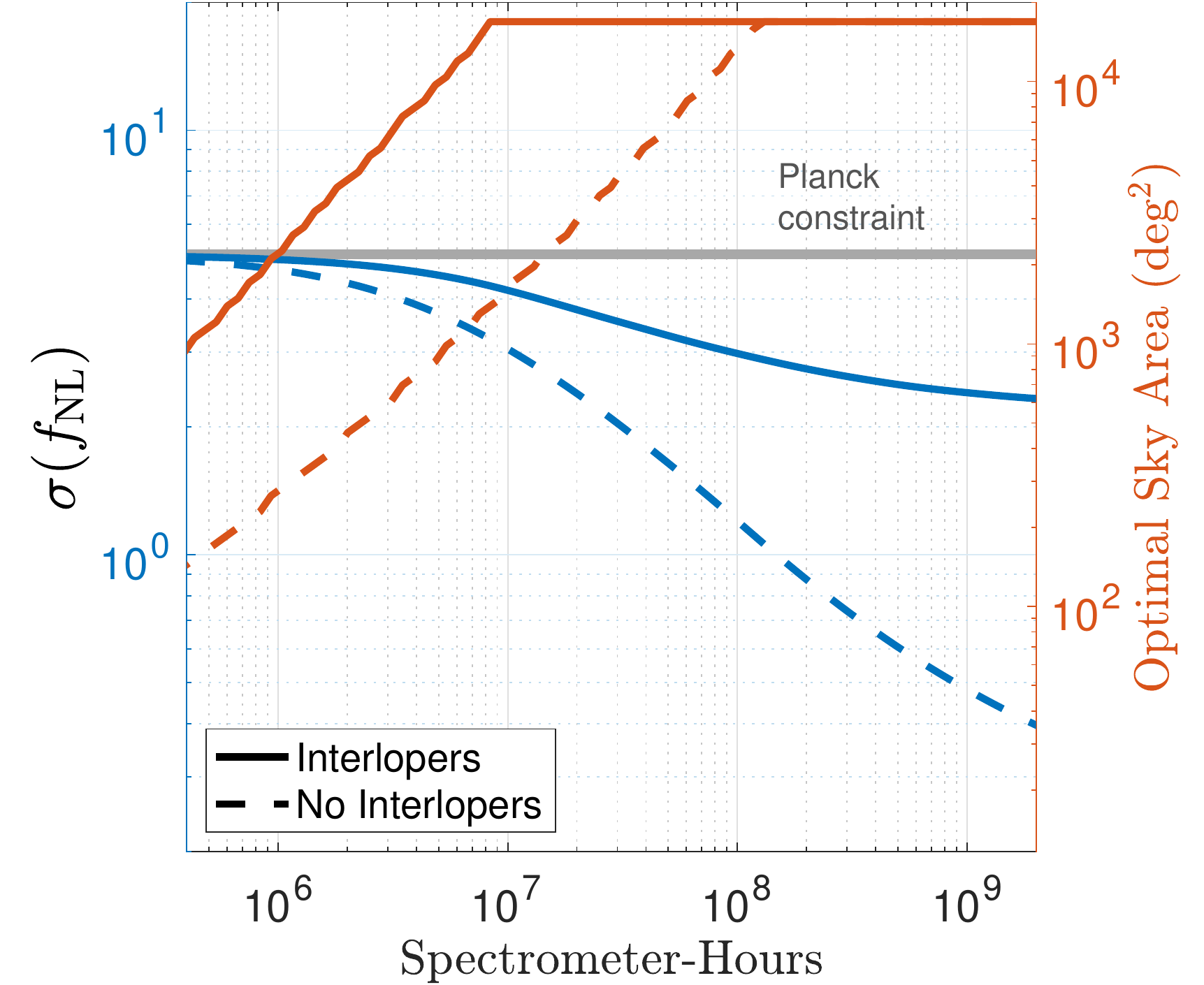}
    \includegraphics[width=0.45 \textwidth]{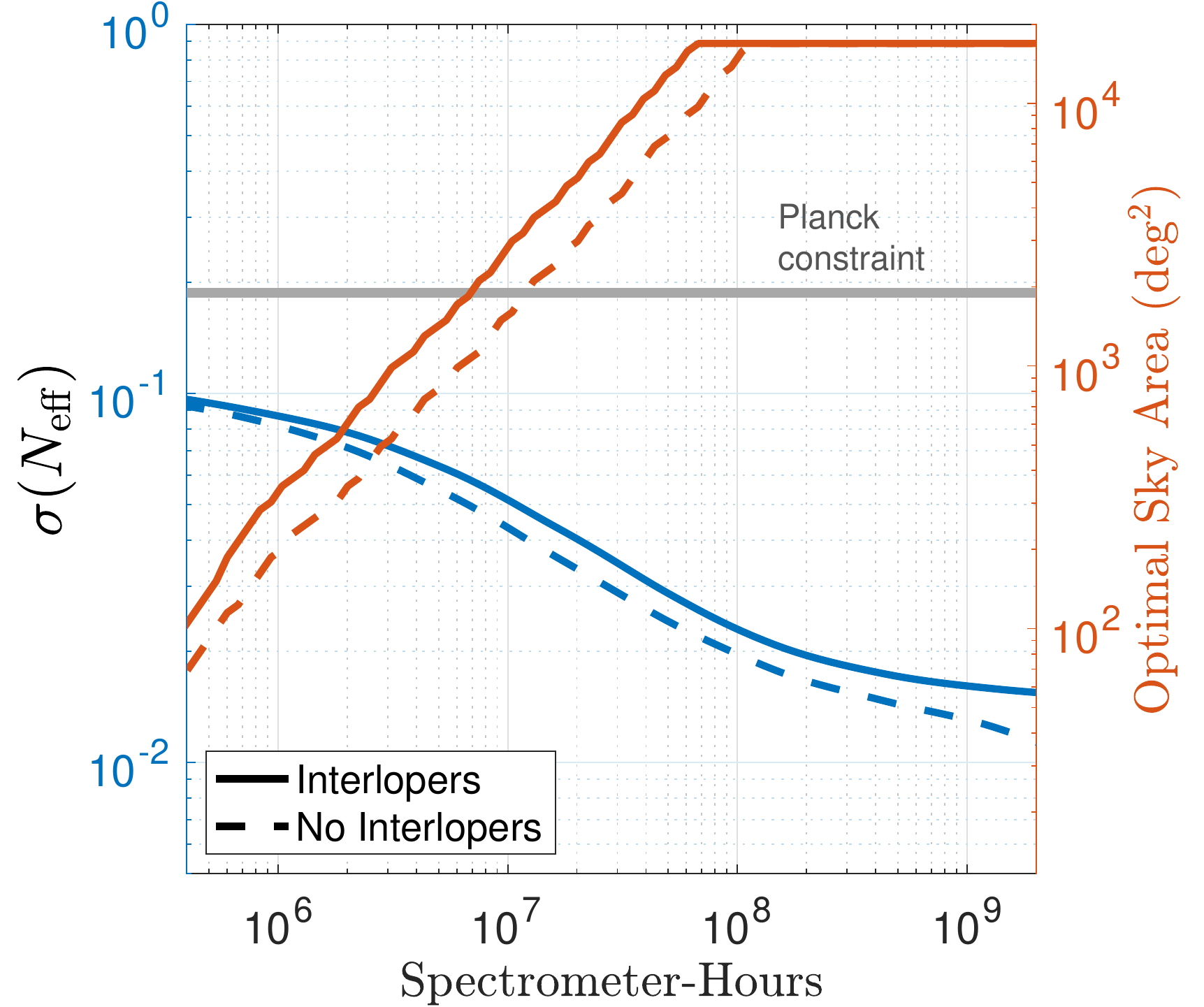}
    \includegraphics[width=0.45 \textwidth]{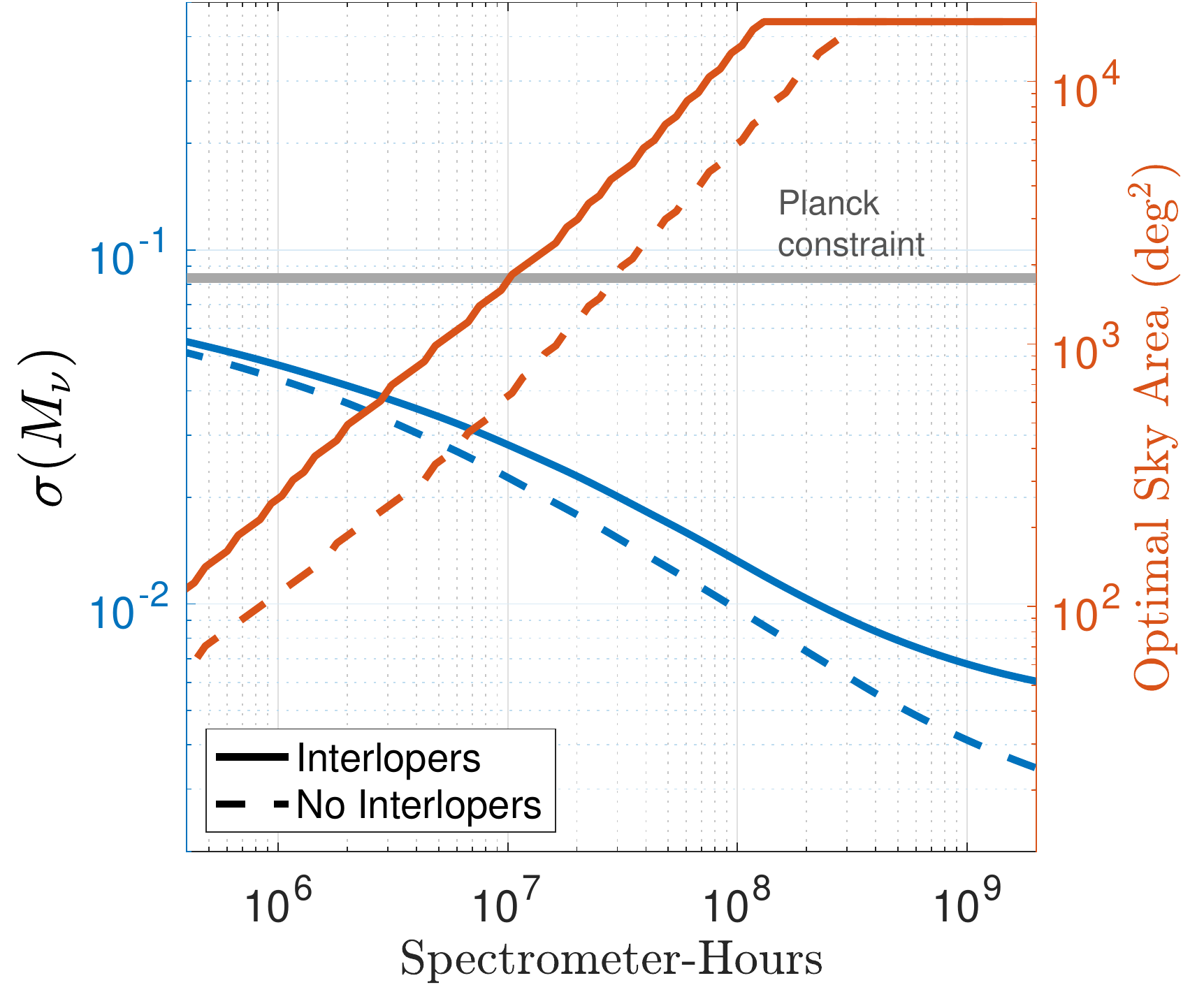}
    \includegraphics[width=0.45 \textwidth]{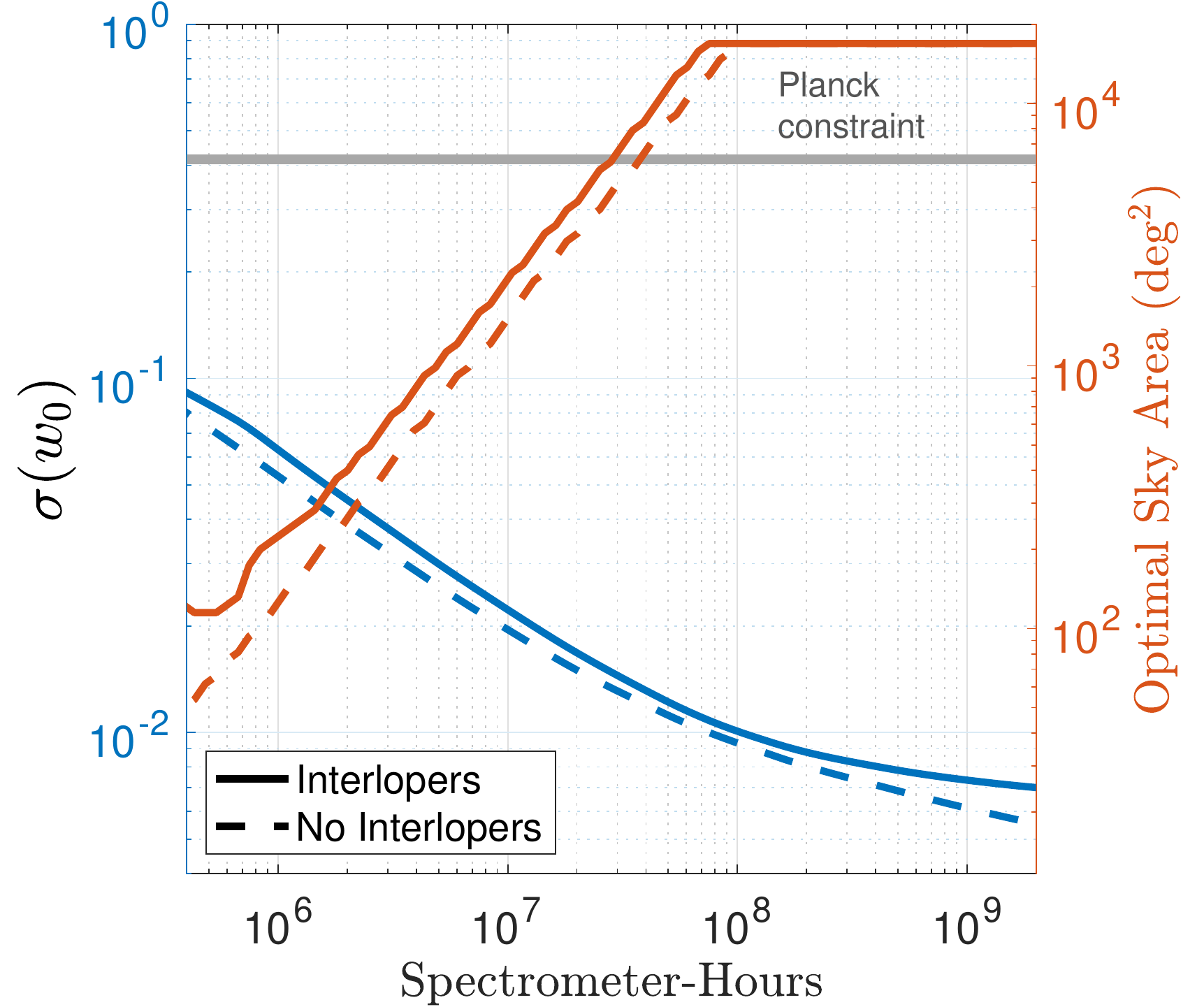}
    \includegraphics[width=0.45 \textwidth]{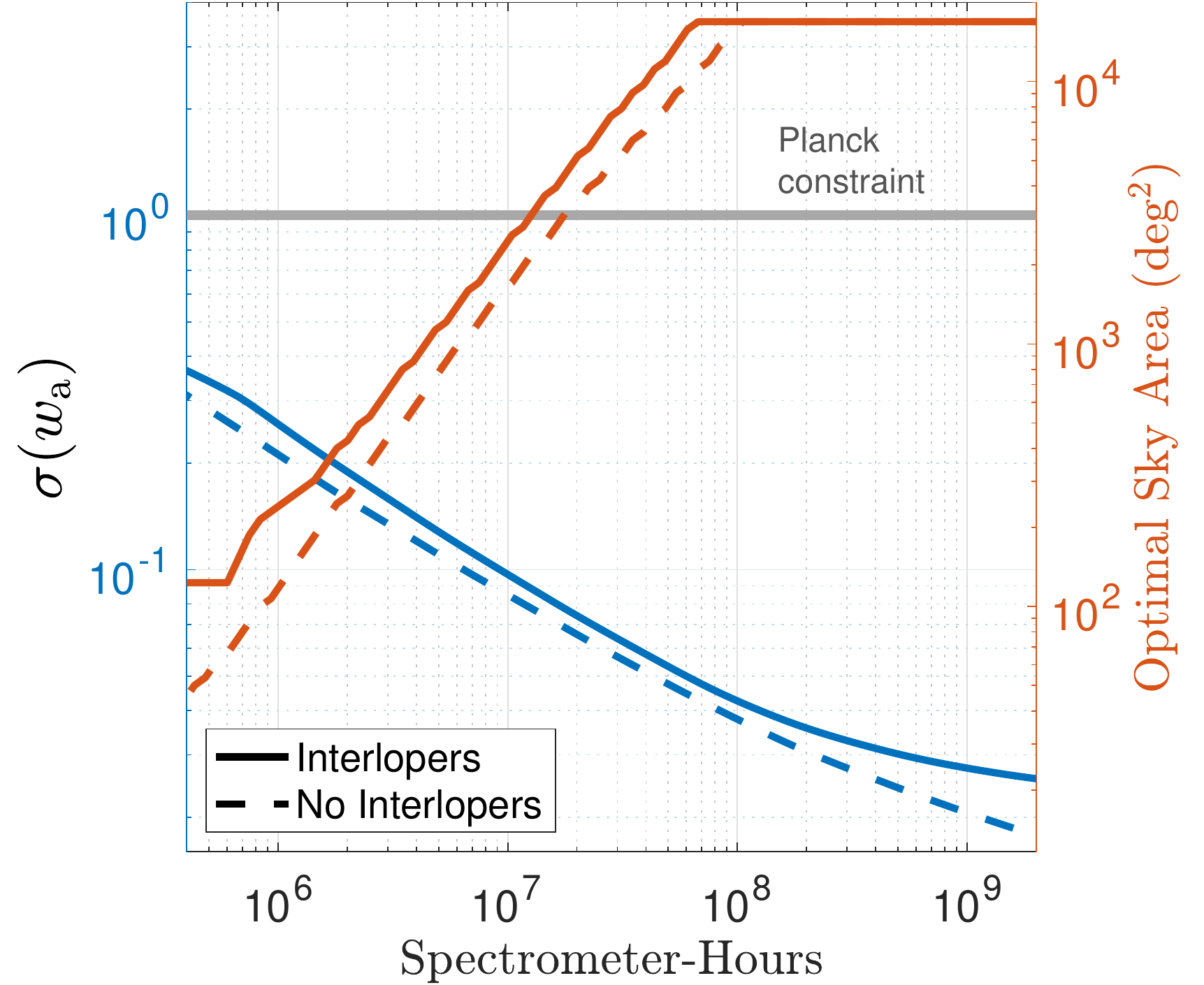}
    \caption{Projected 1$\sigma$ constraints on several extensions to $\Lambda$CDM (blue) and optimal sky coverage to achieve these constraints (red), as a function of a mm-wave LIM experiment's sensitivity parametrized by spectrometer-hours. A spectrometer is assumed to measure 80--310 GHz with $R=300$ from excellent mm-wave observing sites. The dashed lines correspond to the case where interloper lines are neglected (i.e., perfect line separation), while the solid lines include the interlopers as noise terms. In all cases $\Lambda$CDM parameters are marginalized over and Planck priors are imposed.
    }
    \label{fig:1sigma}
\end{figure}

Figure~\ref{fig:1sigma} shows projections for one-parameter extensions to $\Lambda$CDM varying $f_{\rm NL}$, $N_{\rm eff}$, and $M_{\nu}$, and for a three-parameter extension varying $w_0$, $w_a$, and $M_{\nu}$, in blue. The sky fraction that minimizes parameter uncertainties at each value of spectrometer-hours is shown in red, and limits from Planck are shown with horizontal gray lines. In all cases, significant improvements over current  constraints are possible; the projected uncertainties generally plateau after $\sim 10^8-10^9$ spectrometer-hours, when the observable sky fraction hits a maximum of $f_{\rm sky} = 0.4$. In all cases except $f_{\rm NL}^{\rm loc}$, the ``interloper'' worst-case scenario is not significantly less constraining than the best-case scenario\footnote{Without a Planck prior, the impact of interlopers is more relevant \cite{MoradinezhadDizgah:2021upg}.}. The degradation of the $f_{\rm NL}^{\rm loc}$ constraint agrees with the findings of Ref.~\cite{Chen:2021ykb}, which showed that while interloper contamination increases the uncertainty and biases the best-fit value of $f_{\rm NL}^{\rm loc}$, using cross-correlations between lines of interest and interlopers can largely remove the contamination. 

\textbf{Our forecasts indicate that future mm-wave LIM surveys can deliver cosmological constraints that are comparable to or exceed contemporary and planned CMB/LSS experiments.}
LIM experiments in the $\sim 10^7$ spectrometer-hour range can deliver constraints from the clustering power spectrum that are competitive with those from the upcoming generation of spectroscopic galaxy surveys \cite{Font-Ribera:2013rwa}, such as EUCLID \cite{Amendola:2016saw} and DESI \cite{DESI:2016fyo}. Experiments with $\sim 10^8$ spectrometer-hours would improve constraints on most parameters by an additional factor of 2--4. For $N_{\rm eff}$, such a survey would provide comparable constraints to baseline forecasts for CMB-S4 \cite{CMB-S4:2016ple}, while for $M_\nu$, a $\sim 10^7$ spectrometer-hour survey gives similar constraining power to CMB-S4. Projected constraints on $f_{\rm NL}^{\rm loc}$ surpass those from the proposed PUMA 21 cm experiment at $10^7$--$10^8$ spectrometer-hours depending on the degree of interloper mitigation \cite{Karagiannis:2019jjx}
; we also note that our forecasts for PNG are based only on the power spectrum and that significant additional constraining power will come from the bispectrum.

Projections for several of the science targets outlined in Section~\ref{sec:science} rely on effects observable in the DM fluctuations in the linear regime, i.e., the initial conditions of structure formation. Therefore, the number of measurable linear (or weakly nonlinear) modes is a useful metric for comparing the potential of different LSS surveys. The decorrelation of the observed fluctuations from their initial conditions due to nonlinear evolution and instrumental noise is accounted for in defining the number of modes. As in Ref.~\cite{Snowmass2021:largeN}, we compute the {\it primordial physics Figure of Merit (FoM)} to characterize the ``effective'' number of observed modes correlated with the initial conditions (see Appendix \ref{app:FoM} for further details). Sensitivity to several parameters of interest---such as primordial features, local-shape PNG, and EDE---scale as $1/\sqrt{{\rm FoM}}$. 
Ambitious future LIM surveys could achieve a FoM up to 170, with the exact value depending on how well interloper lines can be suppressed. For comparison, the expected FoM for DESI is 0.16, while PUMA expects 26--85 depending on foreground mitigation; see \cite{Snowmass2021:largeN} for more extended comparison across various LSS surveys. 

When interpreting the forecasted parameter constraints and FoM, it is important to note that the LIM technique is still in an early stage of development compared to the CMB and galaxy surveys, and several major caveats exist. First, while we assume perfect knowledge of the nuisance astrophysical parameters (i.e., the mean brightness temperature and line bias, which are degenerate with cosmological parameters), their current uncertainties are large. We anticipate that taking cross-correlations, both internally in multi-line intensity maps and with other probes such as galaxy surveys and CMB lensing, will significantly reduce these uncertainties \cite{obuljen2018}. Second, we assume linear theory for the line power spectrum; marginalizing over nonlinear parameters could degrade constraints, with the of impact depending on the choice of small-scale cutoff \cite{Sailer:2021yzm}. Finally, we assume that large angular and line-of-sight scales can be recovered without a significant $1/f$ component above the $k_{\rm min}$ cutoffs. The degree to which this is possible is uncertain and depends on the scan strategy and atmospheric noise---techniques to recover large scales will be tested on pathfinder experiments.

Although our forecasts are performed specifically for mm-wave measurements from 80--310 GHz, we also anticipate and advocate for measurements of rest-frame far-IR lines outside of this range. Generally, cm-wave or THz measurements employ technology and observation strategies similar to mm-wave, and will improve the statistical power and redshift range of LIM measurements. Ground-based cm-wave surveys of low-J CO lines with $10^7$--$10^8$ spectrometer-hours would complement mm-wave measurements by independently providing percent-level determination of the high-redshift expansion history \cite{bernal2019a,bernal2019b,karkare2018} and PNG constraints from multiple tracers \cite{MoradinezhadDizgah:2018zrs,MoradinezhadDizgah:2018lac,liu_breysse_2021}. Since CO(1-0) and CO(2-1) at $z\gtrsim2$ have no interloper lines to contend with, these measurements will also aid in attributing mm-wave signals to specific lines. Surveys of [CII], [NII], [OI], [OIII], and other lines extending to higher frequencies should enable similar synergies.

\subsection{Experimental program}\label{sec:stages}

We envision a staged experimental approach in which increasing numbers of spectrometers are deployed on ground-based mm-wave telescopes to achieve the science goals outlined in the previous section. This strategy follows the approach used to develop similar technology for current mm-wave observatories measuring the CMB. It offers a natural path to increasing technical readiness and scale, and by the mid-2030s, this program could position mm-wave LIM as a successor to current CMB surveys.

Table~1 outlines possible experimental stages and forecasts for the science cases in Section~\ref{sec:science}, in addition to the primordial physics Figure of Merit. Each stage corresponds to a different experimental scale and is differentiated by the instrumentation and platform used---while current experiments employ a wide array of spectrometer technologies, we anticipate that future projects will converge on high-density on-chip spectrometers deployed at existing CMB facilities. Once the transition to CMB-style instruments is made, we assume that densely-packed spectrometer wafers will be deployed in receivers with multiple ``optics tubes,'' as envisioned for SO, SPT-3G+, and CMB-S4. An optics tube is assumed to hold an array of 400 on-chip spectrometers on a 6'' wafer in the same formfactor as the SO-LAT optics tubes; note that this requires a significant increase in packing density over current on-chip spectrometers (see Section~\ref{sec:invest}).

\begin{table}[t]
\center
\resizebox{\textwidth}{!}{\begin{tabular}{|l|l|l|l|l|l|l|l|l|}
\hline
\makecell{Spec-\\ hrs} & \makecell{Example} & \makecell{Time-\\ scale} & \makecell{$\sigma(f_{\rm NL})$ } & \makecell{$\sigma(M_{\nu})$\\ (meV) }  & \makecell{$\sigma(N_{\rm eff})$} &  \makecell{$\sigma(w_0) \times 10^2$} & \makecell{$\sigma(w_{\rm a}) \times 10^{2}$} & \makecell{FoM} \\ \hline
$10^5$ &  \makecell{TIME, CCAT-p, \\ SPT-SLIM} & 2022 &  5.1 (5.1)  &  61 (65)   &  0.1 (0.11)    & 13 (14)     & 51 (52)   & 0.0015 \\ \hline
$10^6$ &  \makecell{TIME-EXT }                 & 2025 &  4.7 (5)    &  43 (47)   &  0.082 (0.087) & 5.3 (6.3)   & 21 (26)   & (0.09-0.1) \\ [1ex]\hline
$10^7$ &  \makecell{SPT-like\\ 1 tube}         & 2028 &  3.1 (4.2)  &  23 (28)   &  0.043 (0.051) &   2 (2.2)   & 8.5 (9.7) & (1.7-3.1) \\  \hline
$10^8$ &  \makecell{SPT-like\\ 7 tubes}        & 2031 &  1.2 (3)    &  9.7 (13)  &  0.02 (0.023)  & 0.93 (1)    & 3.8 (4.3) & (9.5-28)\\  \hline
$10^9$ &  \makecell{CMB-S4-like\\ 85 tubes}    & 2037 &  0.48 (2.4) &  4.1 (6.8) &  0.013 (0.016) & 0.61 (0.73) & 2.1 (2.8) & (21-108) \\
\hline\hline
\multicolumn{3}{|c|}{Planck} & 5.1 & 83 & 0.187 & 41 & 100 & --- \\
\hline
\end{tabular}}
\label{table:stages}
\caption{Stages of future mm-wave LIM experiments and projected constraints for cosmological science cases, with optimistic (pessimistic) assumptions about the ability to suppress the impact of interloper emission. For each stage we provide an approximate example of the class of instrument required for such a survey. The timescale is a rough estimate of when such a survey could begin operations. All forecasts include Planck priors.}
\end{table}

While this program outlines an ambitious scaling up from current-generation experiments, we note that over the last 10 years similar progress has been made with CMB experiments. Moreover, this progression is feasible using existing 5--10m class CMB facilities, which are well-suited to the degree-to-arcminute scales necessary for these LIM science cases. Experimental techniques for extracting power spectra from extended, low surface brightness emission are mature and large classes of instrumental systematics relevant to the measurement are well-understood and quantified.

\section{Investments Needed  } 
\label{sec:invest}

\subsection{Technology Developments}

While first-generation mm-wave LIM instruments are now being demonstrated, several advances are still needed to enable the sensitivity required to cross the aforementioned science thresholds.  The following hardware-related investments in detectors, readout, and facilities would leverage the considerable investment made in CMB experiments over the last few decades:

\begin{itemize}

\item \textbf{On-Chip Spectrometer Packing Density}: Current approaches to mm-wave spectroscopy (diffraction gratings, Fourier Transform or Fabry-Perot spectroscopy, heterodyne detection) are difficult to scale to large spectrometer counts. \textit{On-chip} spectroscopy, in which the detector and spectrometer are integrated on a silicon wafer, offers a natural path to maximizing the sensitivity of mm-wave receivers. However, while prototype on-chip spectrometers are now being demonstrated \cite{karkare2020, endo2019}, their spatial packing density is still significantly lower than CMB focal planes (primarily due to the physical extent of the spectrometer on the wafer). Innovation in focal plane geometry and layout will be key to enabling high-density close-packed arrays. Our forecasts assume that a 6'' diameter wafer holds $\sim 400$ spectrometers, which fully samples the available focal plane area.

\item \textbf{On-Chip Spectrometer Spectral Resolution}: On-chip spectrometers have demonstrated spectral resolution of $R \sim 300 - 500$ \cite{karkare2020, taniguchi2021}. Improving this to $R \sim 1000$ would move LIM experiments into the regime of spectroscopic surveys. Higher $R$  increases the number of available modes and significantly improves the measurement of redshift-space distortions \cite{chung2019,Sailer:2021yzm}. When mitigating astrophysical systematics, higher $R$ enables higher-fidelity line separation, either by allowing finer application of anisotropic projection effects \cite{Lidz:2016,Cheng:2016} or phase-space templates \cite{Kogut2015,Cheng:2020}, or by requiring less volume to be excised when masking foreground lines \cite{Silva:2014,Breysse:2015,Sun:2018}. Increasing the spectral resolution of on-chip spectrometers requires development of new low-loss dielectric materials.

\item \textbf{Multiplexed Readout}: Spectroscopic pixels require significantly more detectors than their broadband counterparts; a 1000-spectrometer array with $R \sim 300$ would have a detector count approaching that of the entire CMB-S4 experiment. Readout development will be essential to the success of future spectroscopic instruments. New technologies based on state-of-the-art FPGA platforms, such as the RF system-on-chip, promise to dramatically reduce overall readout costs to \$1--2 per channel \cite{lowe2020}.

\item \textbf{Telescopes and Facilities}:
The science goals outlined above can all be accomplished with 5--10 m class telescopes, with unobstructed optics optimized for measurements of extended, low surface brightness emission and minimal instrumental systematics. Several such facilities exist or will be built in the next decade, including SPT, ACT, SO, CCAT-p, and CMB-S4 \cite{SO2019, CCAT2021, cmbs4_DSR}. Reusing these telescopes for the LIM measurement represents a natural step that would extend their lifetime in service of cosmological science similar to CMB.

\end{itemize}

\subsection{Analysis Techniques}
Controlling astrophysical, instrumental, and modeling systematics will also be critical for LIM cosmology. Investments in analysis techniques, which can be validated by pathfinder experiments, will enable future surveys to take advantage of the sensitivity enabled by hardware advances: 

\begin{itemize}
    \item \textbf{Astrophysical systematics}: Both astrophysical continuum and spectral line emission induce systematics in the LIM measurement. Continuum foregrounds/backgrounds can be orders of magnitudes stronger than the signal of interest \cite{Switzer:2019}. However, they generally vary smoothly with frequency, affecting modes at low spectral wavenumber (i.e., where $k_{\rm z}\sim0$). Furthermore, mm-wave continuum emission has been extensively studied at the angular scales of interest by CMB experiments. Spectral line sources---typically referred to as interlopers---are another potential source of systematics. This is particularly true at mm-wave, where CO rotational transitions from a range of redshifts could be present at any given frequency, along with other potential lines \cite{Righi:2008,Breysse:2015}. Mitigating interloper lines remains an active area of research with several methods under investigation, including masking \cite{Silva:2014,Breysse:2015,Sun:2018}, leveraging differential effects in projection from angular/spectral dimensions to physical space (i.e., the Alcock-Paczy\'{n}ski effect \cite{Alcock1979,Lidz:2016,Cheng:2016}), coherently removing power  \cite{Kogut2015,Cheng:2020}, and using machine learning techniques \cite{moriwaki2020, moriwaki2021}. Preliminary work suggests that these methods should be highly effective in suppressing the effects and biases of interloper emission, but more effort is needed, along with validation on data from pathfinder experiments. 
    \item \textbf{Instrumental systematics}: Instrumental systematics can arise from a variety of mechanisms, including calibration errors, detector gain variations, and reflections in optics. Of particular concern are effects that induce frequency-dependent structure on otherwise-smooth spectral response \cite{Keating2015,Barry:2016}. At millimeter wavelengths the atmosphere presents an additional challenge for ground-based experiments, since the telluric lines of ozone, water, oxygen, and other chemical species could imprint frequency-dependent structure in the data. Although these systematics are both site- and instrument-specific, mm-wave LIM will benefit from observational techniques developed by CMB experiments, in addition to strategies tested on pathfinder instruments \cite{Keating2016,Foss2021}.
    \item \textbf{Modeling systematics:} Realizing LIM's full potential requires accurate theoretical models of the signal, matching the measurement sensitivity. Like galaxy redshift surveys, this requires both state-of-the art numerical simulations and analytical tools, the latter usually validated by the former. Developments on both fronts are needed for LIM observables. On the analytic side, existing models based on the halo-model framework \cite{Seljak:2000gq,Cooray:2002dia} require two ingredients:\ relating the line intensity to DM halos, and relating the halo properties to the underlying DM distribution. While the latter can be modeled accurately using perturbative models of LSS, there are large uncertainties on the former, primarily driven by lack of observational constraints \cite{MoradinezhadDizgah:2021dei,Carrasco:2012cv,Assassi:2014fva,Senatore:2014via}. Ongoing and future LIM experiments \cite{Crites2014,Lagache_2018,Aravena:2019tye,cooray2019cdim,Essinger_Hileman_2020,Cleary:2021dsp} will inform this poorly-constrained theory space. The combination of observational data and small-volume simulations of hydrodynamical galaxy formation and radiative transfer can be used to construct ``effective'' models of line luminosity, which are simple enough to be incorporated into semi-analytic simulations (based on N-body simulations) over cosmological volumes, and into analysis of future LIM data \cite{Leung2020,Yang:2020lpg,Kannan:2021ucy}. A consistent theoretical framework for the line emission is needed to fully take advantage of multi-line observational capability \cite{Sun2019}.  
\end{itemize}

\section{Conclusions} 
\textbf{Line intensity mapping at millimeter wavelengths is a powerful tool for testing fundamental physics and cosmology by measuring large-scale structure over large volumes and at high redshift}. Far-IR emission lines such as CO and [CII] trace galaxies through most of the history of the Universe, and are redshifted to the mm-wave atmospheric windows where they can be detected from the ground. Observations from proven observing sites, mapping out the Universe from $0 < z <  10$, could measure an order of magnitude more primordial modes than contemporary galaxy surveys and provide transformational constraints on numerous cosmological science cases, including inflation, dark energy, dark matter, and neutrinos and relativistic species.
\textbf{To develop this new observational technique, we recommend support for mm-wave LIM projects as they transition from pathfinders to experiments with significant cosmological constraining power.} This transition will be enabled by investments in the following technical areas:
\begin{itemize}
    \item \textbf{Instrumentation}: Development in mm-wave detectors, spectrometers, and readout will significantly increase instantaneous sensitivity.
    \item \textbf{Analysis}: Development of techniques to mitigate astrophysical, instrumental, and modeling systematics will enable robust extraction of the cosmological signal.
\end{itemize}
All of these efforts leverage decades of investment and experience from CMB and LSS experiments in detector technology, facilities, observing strategy, and analysis.

\section{Acknowledgements}

We thank Adam Anderson, Jos\'e Luis Bernal, Matt Bradford, Clarence Chang, Jacques Delabrouille, Dan Marrone, Gabriela Sato-Polito, Sara Simon, and Michael Zemcov for helpful comments.

\appendix
\section{Primordial Physics Figure of Merit} \label{app:FoM}

To quantify the information content of the LIM power spectrum in constraining primordial physics, we calculate the primordial physics FoM as defined in \cite{Snowmass2021:largeN}. We compute the linear matter power spectrum using the \url{CLASS} Boltzmann code\footnote{\url{https://github.com/lesgourg/class_public}} \cite{Blas:2011rf}, and model the nonlinear line power spectrum using redshift-space Eulerian perturbation theory at one-loop level, including the effective field-theory terms and infrared resummation, using the \url{velocileptors} package \footnote{\url{https://github.com/sfschen/velocileptors}} \cite{Chen:2020fxs}. The line biases are computed by assuming Sheth-Tormen halo biases and co-evolution predictions as in \cite{MoradinezhadDizgah:2021dei}. We set the small-scale cut ($k_{\rm max}$) according to instrument resolutions in the angular and line-of-sight directions, and further impose the cutoff of $k_{\rm max}<1 \ h {\rm  Mpc}^{-1}$. The same large-scale cut is set as in the parameter forecasts. Instrumental noise is computed as in \cite{MoradinezhadDizgah:2021upg}, and interlopers are accounted for as additional noise terms. 
\begin{figure}[h]
    \centering
    \includegraphics[width=0.6 \textwidth]{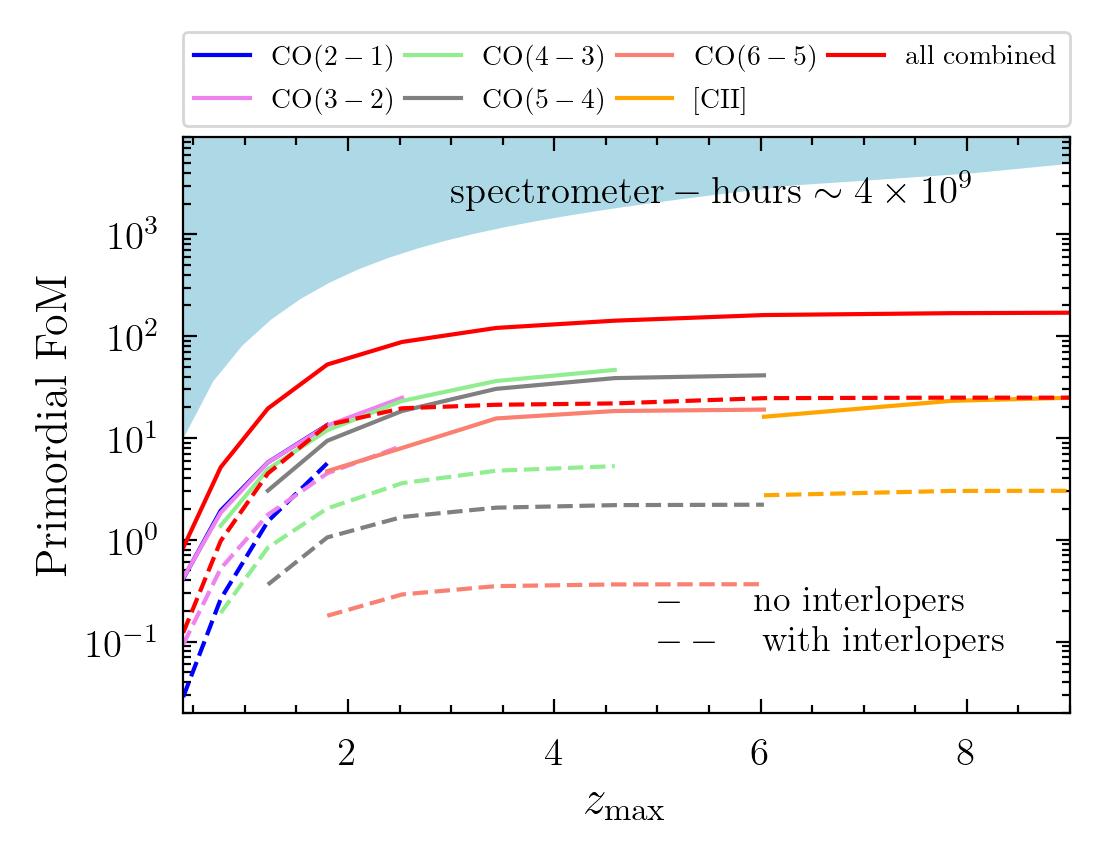}\vspace{-.28in}
    \caption{Primordial physics FoM as a function of maximum observed redshift, for a survey with $\sim 4 \times 10^9$ spectrometer-hours. The shaded blue region is excluded by cosmic variance for a full-sky survey probing all six lines. The solid (dashed) lines assume best- (worst-) scenario of perfect (no) cleaning of interlopers. The red lines show the combined FoM.}
    \label{fig:FoM}
\end{figure}
Figure~\ref{fig:FoM} shows the FoM as a function of maximum observed redshift for $\sim 4 \times 10^9$ spectrometer-hours, for each individual line and their combination. The solid and dashed lines correspond to including or neglecting the interloper lines in the noise. The blue shaded region is the cosmic variance limit for a full-sky survey. 
It is important to recognize that an experiment's constraining power cannot be fully condensed in a single FoM. For instance, while the total FoM (in red) appears to saturate at high redshift, access to the high-redshift regime is particularly powerful in breaking parameter degeneracies, as illustrated by our parameter forecasts. \vspace{-.1in}

\bibliographystyle{unsrturltrunc6.bst}
\bibliography{refs}

\end{document}